\documentclass[review, 1p]{elsarticle}

\usepackage{amsmath,amssymb,bm}
\usepackage{lineno,hyperref}

\journal{Physics Letters B}

\begin{document}

\begin{frontmatter}

\title{Multiple Chirality in Nuclear Rotation: A Microscopic View}

\author[ANL]{P. W. Zhao
\corref{mycorrespondingauthor}
}
\cortext[mycorrespondingauthor]{Corresponding author}
\ead{pwzhao@pku.edu.cn}

\address[ANL]{Physics Division, Argonne National Laboratory, Argonne, Illinois 60439, USA}

\begin{abstract}
Covariant density functional theory and three-dimensional tilted axis cranking are used to investigate multiple chirality in nuclear rotation for the first time in a fully self-consistent and microscopic way. 
Two distinct sets of chiral solutions with negative and positive parities, respectively, are found in the nucleus $^{106}$Rh. 
The negative-parity solutions reproduce well the corresponding experimental spectrum as well as the $B(M1)/B(E2)$ ratios of the transition strengths. 
This indicates that a predicted positive-parity chiral band should also exist. 
Therefore, it provides a further strong hint that multiple chirality is realized in nuclei.
\end{abstract}


\end{frontmatter}


\section{\label{sec1} Introduction}

\begin{figure}[!htbp]
\centering
\includegraphics[width=5cm]{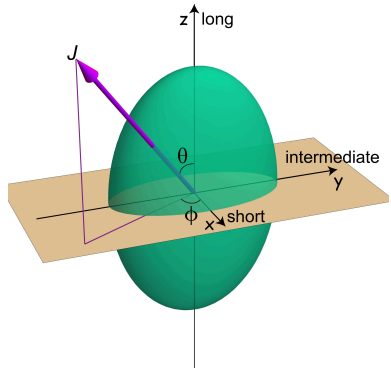}
\caption{(color online). A schematic picture for the aplanar rotation of a triaxial nucleus. The arrow $\bm{J}$ denotes the total angular momentum vector, and the short, intermediate, and long axes are denoted by $x$, $y$, and $z$, respectively.}
\label{fig1}
\end{figure}

Chirality is a well-known phenomenon in many fields, such as chemistry, biology, molecular and particle physics, etc. 
In nuclear physics, chirality was originally suggested by Frauendorf and Meng in 1997~\cite{Frauendorf1997Nucl.Phys.A131}. 
It represents a novel feature of triaxial nuclei rotating around an axis which lies outside the three planes spanned by the  principal axes of the triaxial ellipsoidal density distribution; i.e., {\it aplanar rotation}. 
As depicted in Fig.~\ref{fig1}, the short, intermediate, and long principal axes of a triaxial nucleus form a screw with respect to the angular momentum vector $\bm{J}$; resulting in left- and right-handed systems, which correspond to the same magnitude of the angle $\phi$ but with opposite signs. 
The two chiral systems differ from each other by their intrinsic chirality, and are thus related by the chiral operator $TR(\pi)$ that combines time reversal $T$ and spatial rotation by $\pi$. 

The broken chiral symmetry in the body-fixed frame should be restored in the laboratory frame. 
This gives rise to the so-called chiral doublet bands, which consist of a pair of nearly-degenerate $\Delta I=1$ sequences with the same parity (for reviews see Refs.~\cite{Frauendorf2001Rev.Mod.Phys.463,Meng2010JPhysG.37.64025}).  
The chiral geometry of a rotating triaxial nucleus requires substantial angular momentum components along all of the three principal axes. 
For actual nuclear systems, this relates to configurations where high-$j$ valence particles and holes align along the short and long axes, respectively, and the collective core rotates around the intermediate axis. 
Based on such configurations, so far, many candidates for chiral doublet bands have been reported experimentally in the $A\sim 80$, 100, 130, and 190 mass regions of the nuclear chart; see e.g., Refs.~\cite{Starosta2001Phys.Rev.Lett.971,Zhu2003Phys.Rev.Lett.132501,Vaman2004Phys.Rev.Lett.32501,Joshi2004Phys.Lett.B135,Alcantara-Nunez2004Phys.Rev.C24317,Timar2004Phys.Lett.B178,Grodner2006Phys.Rev.Lett.172501,Joshi2007Phys.Rev.Lett.102501,Mukhopadhyay2007Phys.Rev.Lett.172501,Grodner2011Phys.Lett.B46,Wang2011Phys.Lett.B40,Ayangeakaa2013Phys.Rev.Lett.172504,Kuti2014Phys.Rev.Lett.32501,Tonev2014Phys.Rev.Lett.52501,Liu2016Phys.Rev.Lett.112501}.

These observations indicate that chirality is not restricted to a specific configuration in a certain mass region. 
In particular, two pairs of chiral bands with different configurations were discussed respectively in Refs.~\cite{Alcantara-Nunez2004Phys.Rev.C24317,Timar2004Phys.Lett.B178} for $^{105}$Rh. 
The resultant possibility of having more than one pair of chiral bands in a single nucleus was demonstrated by searching for triaxial chiral configurations in Rh isotopes using constrained relativistic mean-field calculations in Ref.~\cite{Meng2006Phys.Rev.C37303}, which introduced the acronym M$\chi $D for multiple chiral doublet bands. 
Further studies were carried out in Refs.~\cite{Peng2008Phys.Rev.C24309,Yao2009Phys.Rev.C67302,Li2011Phys.Rev.C37301}. 
Strong experimental evidence for M$\chi$D bands has been reported in $^{133}$Ce~\cite{Ayangeakaa2013Phys.Rev.Lett.172504} and $^{103}$Rh~\cite{Kuti2014Phys.Rev.Lett.32501}. Recently, octupole correlations between M$\chi$D bands in $^{78}$Br have also been observed~\cite{Liu2016Phys.Rev.Lett.112501}.

On the theoretical side, the triaxial particle-rotor model (PRM)~\cite{Frauendorf1997Nucl.Phys.A131,Peng2003Phys.Rev.C44324,Koike2004Phys.Rev.Lett.172502,Zhang2007Phys.Rev.C44307,Qi2009Phys.Lett.B175} has been extensively used in studies of chiral doublet bands. 
However, such analyses are all phenomenological and are fitted to the data in one way or another. 
The tilted axis cranking approach allows one to study nuclear chiral rotation based on a microscopic mean field. 
For nuclear chirality, a three-dimensional tilted axis cranking (3DTAC) calculation is required~\cite{Frauendorf2001Rev.Mod.Phys.463}, in which the cranking axis could lie outside the three planes spanned by the principal axes; i.e., both $\theta$ and $\phi$ in Fig.~\ref{fig1} could be nonzero.  
The chiral solutions of the 3DTAC approach have been examined based on a Woods-Saxon potential combined with the Shell correction method~\cite{Dimitrov2000Phys.Rev.Lett.5732} or a Skyrme-Hartree-Fock mean field~\cite{Olbratowski2004Phys.Rev.Lett.52501}. 
In particular, the former was also used to interpret the likelihood of two chiral bands with different configurations in the observed band structures of $^{105}$Rh~\cite{Alcantara-Nunez2004Phys.Rev.C24317,Timar2004Phys.Lett.B178}. 

The nuclear density functional theories (DFTs) provide a fully self-consistent mean field which depends entirely on a universal energy density functional based on the effective nuclear interactions and, thus, would provide a further strong test of the multiple chirality in nuclei. 
Moreover, the covariant version of DFT exploits basic properties of QCD at low energies, in particular, the presence of symmetries and the separation of scales~\cite{Lalazissis2004}. It provides a consistent treatment of the spin degrees of freedom, includes the complex interplay between the large Lorentz scalar and vector self-energies induced at the QCD level~\cite{Cohen1992Phys.Rev.C1881}, and naturally provides the nuclear currents induced by the spatial parts of the vector self-energies, which play an essential role in rotating nuclei.

To describe nuclear rotation, covariant DFT has been extended with the cranking method~\cite{Koepf1989Nucl.Phys.A61,Afanasjev1999Phys.Rep.1,Meng2013Front.Phys.55,Meng2016PhysicaScripta53008}. 
In particular, the two-dimensional tilted axis cranking covariant DFT~\cite{Zhao2011Phys.Lett.B181} has been successfully used to describe the magnetic rotation bands~\cite{Zhao2011Phys.Lett.B181,Yu2012Phys.Rev.C24318}, antimagnetic rotation bands~\cite{Zhao2011Phys.Rev.Lett.122501,Zhao2012Phys.Rev.C54310}, transitions of nuclear spin orientation~\cite{Zhao2015Phys.Rev.C34319}, and linear alpha cluster bands~\cite{Zhao2015Phys.Rev.Lett.22501}, and has demonstrated high predictive power~\cite{Meng2013Front.Phys.55,Meng2016PhysicaScripta53008}. 
Therefore, in the present work, we report on the first application of the 3DTAC method based on covariant density functional theory for nuclear chirality.

\section{\label{sec2} Theoretical Framework}
Covariant DFT starts from a Lagrangian, and the corresponding Kohn-Sham equations have the form of a Dirac equation with effective fields $S(\bm{r})$ and $V^\mu(\bm{r})$ derived from this Lagrangian~\cite{Ring1996Prog.Part.Nucl.Phys.193,Vretenar2005Phys.Rep.101,Meng2006Prog.Part.Nucl.Phys.470,Niksic2011Prog.Part.Nucl.Phys.519,Meng2015}. In the 3DTAC method, these fields are triaxially deformed, and the calculations are carried out in the intrinsic frame rotating with a constant angular velocity vector $\bm{\omega}$, pointing in an arbitrary direction in space:
\begin{equation}\label{Eq.Dirac}
	\left[\bm{\alpha}\cdot(\bm{p}-\bm{V}) + \beta(m+S) + V - \bm{\omega}\cdot\hat{\bm{J}} \right]\psi_k = \epsilon_k\psi_k.
\end{equation}
Here, $\hat{\bm{J}}$ is the total angular momentum of the nucleon spinors, and the fields $S$ and $V^\mu$ are connected in a self-consistent way to the nucleon densities and current distributions, which are obtained from the single-nucleon spinors $\psi_k$~\cite{Zhao2011Phys.Lett.B181,Zhao2012Phys.Rev.C54310}. 
The iterative solution of these equations yields single-particle energies, expectation values for the three components $\langle \hat{J}_i\rangle$ of the angular momentum, total energies, quadrupole moments, transition probabilities, etc. 
The magnitude of the angular velocity $\bm{\omega}$ is connected to the angular momentum quantum number $I$ by the semiclassical relation $\langle \hat{\bm{J}}\rangle\cdot\langle\hat{\bm{J}} \rangle=I(I+1)$, and its orientation is determined by minimizing the total Routhian self-consistently (see below). 

In this work, the point-coupling Lagrangian PC-PK1~\cite{Zhao2010Phys.Rev.C54319} is adopted, and the calculations are free of additional parameters. 
The present study focuses on chirality in the odd-odd nucleus $^{106}$Rh~\cite{Joshi2004Phys.Lett.B135}, which has been regarded as one of the best known examples for chiral doublet bands~\cite{Meng2010JPhysG.37.64025}. 
The Dirac equation [Eq.~(\ref{Eq.Dirac})] is solved in a three-dimensional Cartesian harmonic oscillator basis with 10 major shells. 

\begin{figure}[!htbp]
\centering
\includegraphics[width=8.cm]{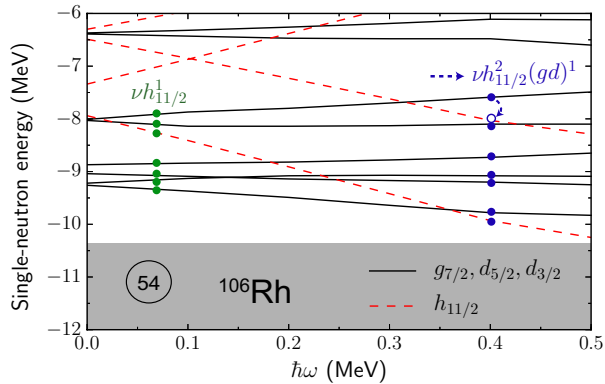}
\caption{(color online) Single-neutron levels near the Fermi surface for $^{106}$Rh as a function of the cranking frequency based on the ground-state configuration. Positive- (negative-) parity states are marked by solid (dashed) lines. The occupied levels are denoted by filled circles. Further details are given in the text.
}
\label{fig2} 
\end{figure}

For $^{106}$Rh, we first solve iteratively Eq.~(\ref{Eq.Dirac}) by placing the nucleons self-consistently in the single-particle orbitals according to their energies starting from the bottom of the well. 
This leads automatically to the ground-state configuration where one neutron particle sits at the bottom of the $h_{11/2}$ shell, and three proton holes are at the top of the $g_{9/2}$ shell, but two of them are anti-aligned. 
In short, we denote this configuration as $\pi g_{9/2}^{-1}\otimes\nu h_{11/2}^{1}$. 
With increasing cranking frequency $\hbar\omega$, the occupation of the protons remains unchanged up to $\hbar\omega = 0.7$ MeV. 
However, as shown in Fig.~\ref{fig2} by the arrow at high frequencies, the last occupied neutron in the $gd$ shell with positive parity can easily drop to the $h_{11/2}$ shell with negative parity. 
Therefore, this provides another configuration labeled as $\pi g_{9/2}^{-1}\otimes\nu h_{11/2}^{2}(gd)^1$ with the two neutrons aligning at the bottom of the $h_{11/2}$ shell. 

Apart from the high-$j$ particles and holes involved in these two configurations, 
it is also found that their deformation parameters $(\beta,\gamma)$ are respectively $(0.25,22.8^\circ)$ for the negative-parity configuration and $(0.27,20.8^\circ)$ for the positive-parity one; reflecting substantial triaxial deformations and, thus, chiral bands could possibly be built on top of both configurations.

In order to examine the possible existence of multiple chirality in $^{106}$Rh based on the two configurations, configuration-fixed 3DTAC calculations have been carried out self-consistently. 
For each configuration, the occupation of the valence neutrons in the $h_{11/2}$ shell are fixed by tracing the single-neutron levels
with increasing frequencies, while other nucleons are treated self-consistently by filling the orbitals according to their energies from the bottom of the well. 

\section{\label{sec3} Results and discussion}

\begin{figure}[!htbp]
\centering
\includegraphics[width=8.cm]{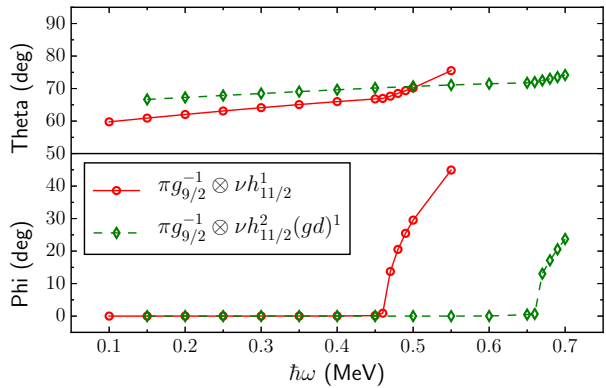}
\caption{(color online) Evolution of the orientation angles $\theta$ and $\phi$ for the angular momentum $\bm{J}$ (defined in Fig.~\ref{fig1}) as driven by the increasing cranking frequency $\omega$ for the two configurations $\pi g_{9/2}^{-1}\otimes\nu h_{11/2}^{1}$ (circles) and $\pi g_{9/2}^{-1}\otimes\nu h_{11/2}^{2}(gd)^1$ (diamonds). 
}
\label{fig3} 
\end{figure}

To examine the possible presence of chiral geometry, it is crucial to check the calculated orientation angles $\theta$ and $\phi$ of the total angular momentum $\bm{J}$ in the intrinsic frame. 
Fig.~\ref{fig3} illustrates the evolution of $\theta$ and $\phi$ driven by the increasing cranking frequency. 
It is clear that they vary in a similar way for both configurations.  
The polar angle $\theta$ is always larger than $45^\circ$ since the angular momentum alignment along the short axis, mainly contributed by the neutron particle(s) in the $h_{11/2}$ shell, is remarkably larger than that along the long axis contributed mainly by the proton hole in the $g_{9/2}$ shell. 
At low frequencies, the azimuth angle $\phi$ vanishes; providing a planar rotation, where the angular momentum lies in the short-long plane (see Fig.~\ref{fig1}). 
Above some finite frequencies, however, the $\phi$ values become nonzero and, thus, the rotation becomes chiral.  
Therefore, a transition between planar and chiral rotation has been found for both configurations, and the corresponding transition points are $\hbar\omega \sim 0.46$ MeV for the $\pi g_{9/2}^{-1}\otimes\nu h_{11/2}^{1}$ configuration and $\sim 0.66$ MeV for the $\pi g_{9/2}^{-1}\otimes\nu h_{11/2}^{2}(gd)^1$ one, respectively. 

It should be noted that TAC gives only the classical orientation, around which the angular momentum $\bm{J}$ can execute a quantal motion. 
In the planar regime ($\phi = 0$), the angular momentum vector $\bm{J}$ oscillates around the planar equilibrium into the left- and right-handed sectors; leading to so-called chiral vibrations~\cite{Starosta2001Phys.Rev.Lett.971}. 
As a result, two separate bands are expected to be observed, corresponding to the first two vibrational states. 

\begin{figure}[!htbp]
\centering
\includegraphics[width=8.cm]{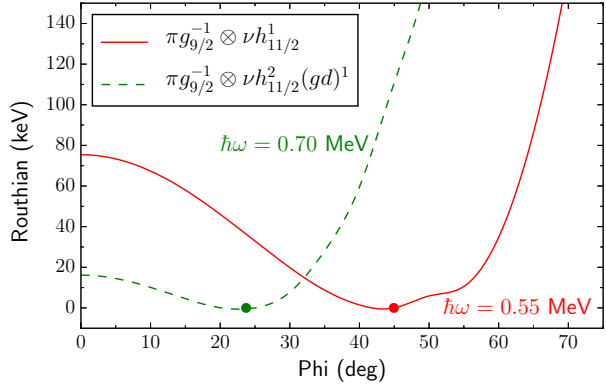}
\caption{(color online) Total Routhian curves as functions of the azimuth angle $\phi_\omega$ of the angular velocity $\bm{\omega}$, which is determined by minimizing the total Routhian with respect to the polar angle $\theta_\omega$ for each given value of $\phi_\omega$.
The two Routhian curves are renormalized to their minima for each configuration, respectively, and the solid circles denote the positions of the minima.}
\label{fig4} 
\end{figure}

In the chiral regime, the energy differences between the chiral twin bands could remain due to tunneling between the left- and right-handed sectors. 
This can be seen from Fig.~\ref{fig4}, where the total Routhian curves are shown as functions of $\phi_\omega$ for both configurations at their largest frequencies.
Here, the azimuth angle $\phi_\omega$ and the polar angle $\theta_\omega$ are used to represent the orientation of the angular velocity $\bm{\omega}$. 
The total Routhian curves are determined by minimizing the total Routhian with respect to $\theta_\omega$ for each given value of $\phi_\omega$. 
It is found that the orientation of $\bm{\omega}$ is parallel to the angular momentum vector $\bm{J}$ at the position of the lowest Routhian for both configurations. 
This indicates that self-consistency has been fully achieved in the present calculations. 

It should be noted that the Routhians at $\pm\phi_\omega$ are degenerate 
and, thus, one can actually find two degenerate minima of the Routhian for each configuration. 
They correspond to the left- and right-handed sectors, respectively. 
As shown in Fig.~\ref{fig4}, the minima of the total Routhian are rather soft in the $\phi_\omega$ direction. 
The barrier of the Routhian at $\phi_\omega=0$ is only several tens of keV in magnitude for both configurations. 
This indicates that tunneling between the left and the right-handed sectors could be substantial, and a strong degeneracy of the chiral twin bands is, thus, not expected.

\begin{figure}[!htbp]
\centering
\includegraphics[width=8cm]{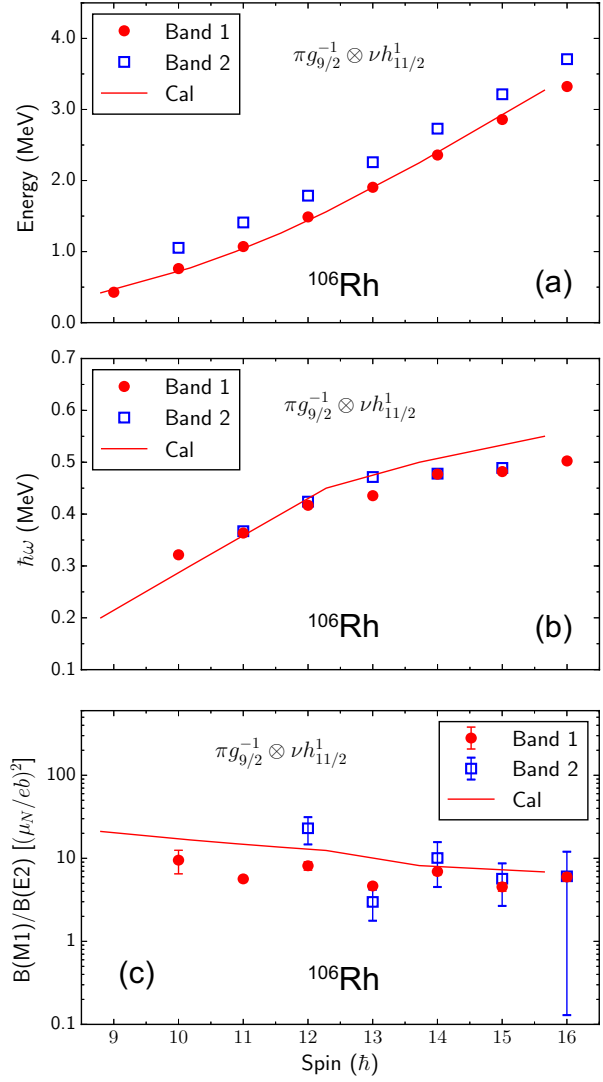}
\caption{(color online) Calculated rotational excitation energies (top), rotational frequencies (middle), and $B(M1)/B(E2)$ ratios for the configuration $\pi g_{9/2}^{-1}\otimes\nu h_{11/2}^{1}$ in $^{106}$Rh, as a function of the angular momentum in comparison with the data for the chiral bands observed in Ref.~\cite{Joshi2004Phys.Lett.B135}. The excitation energies are renormalized to the ground state.}
\label{fig5} 
\end{figure}

A pair of negative-parity chiral bands has been observed in $^{106}$Rh~\cite{Joshi2004Phys.Lett.B135}, and the data on the excitation energies, rotational frequencies, and $B(M1)/B(E2)$ ratios are given in Fig.~\ref{fig5} in comparison with the calculated results, based on the negative-parity configuration $\pi g_{9/2}^{-1}\otimes\nu h_{11/2}^{1}$.  
At the present mean-field level, this does not take into account either the chiral vibrations nor the tunneling between the left- and right-handed sectors. 
Therefore, the energy splitting between the two partner bands cannot be calculated. 
However, it can be clearly seen in Fig.~\ref{fig5}(a) that the experimental excitation energies of the lower band can be reproduced well.
For its partner band, further extensions going beyond the mean field by using, for instance, the methods of the random phase approximation~\cite{Mukhopadhyay2007Phys.Rev.Lett.172501} or the collective Hamiltonian~\cite{Chen2013Phys.Rev.C24314,Chen2016Phys.Rev.C44301} will be required in the framework of DFTs. 

As shown in Fig.~\ref{fig5}(b), the experimental rotational frequencies $\hbar\omega$ for the two partner bands behave very similarly with respect to the angular momentum, and they agree well with the calculated results. 
In particular, the data indicate a kink around $I=12\hbar$, where the slope of the $\hbar\omega$ vs spin curve changes abruptly. 
In comparison with the calculated results, it is found that the appearance of this kink is due to the change from the planar to the chiral solutions around $\hbar\omega\sim 0.46$ MeV. 
In the planar regime, the angular momentum is generated in the plane spanned by the short and long axes, while in the chiral one, it increases along the intermediate axis, which has larger moments of inertia. 
Note that the observed energy separation between the twin bands is almost constant at $\sim 300$ keV, regardless of the planar-chiral transition. 
This is consistent with the rather soft Routhian curve obtained in the chiral regime (Fig.~\ref{fig4}), which indicates a substantial tunneling between the left and right-handed minima, and this may explain the lack of degeneracy in the chiral regime. 

The $B(M1)$ and $B(E2)$ transition probabilities can be calculated in the semiclassical approximation
from the magnetic and quadrupole moments, respectively~\cite{Frauendorf1997Nucl.Phys.A131}. 
Here, the magnetic moments are derived from the relativistic electromagnetic current operator~\cite{Peng2008Phys.Rev.C24313} with  the Dirac effective mass scaled approximately to the nucleon mass by introduction of a factor 0.58 that accounts for the so-called back-flow effects, which have been calculated in infinite nuclear matter by a Ward identity~\cite{Bentz1985Nucl.Phys.A593,Arima2011SciChinaSerG-PhysMechAstron188}. 
It can be seen from Fig.~\ref{fig5}(c) that the calculated $B(M1)/B(E2)$ ratios are in a good agreement with the data~\cite{Joshi2004Phys.Lett.B135}. 

Note that the deformation parameters $(\beta,\gamma)$ change only slightly along the band from $(0.25,22.8^\circ)$ to $(0.23,26.0^\circ)$. 
Moreover, the TAC calculations indicate a slow increase of $\theta$ and $\phi=0$ for the states below $\hbar\omega=0.46$ MeV ($I\sim$12$\hbar$), so the corresponding $B(E2)$ values are almost constant.
However, the $B(M1)$ values decrease smoothly because of the closing of the neutron and proton angular momentum vectors, which mainly align along the short and long axes, respectively. 
This leads to the smooth-decreasing tendency for the $B(M1)/B(E2)$ ratios. 
Above $\hbar\omega=0.46$ MeV, however, the rapid increase of $\phi$ causes the steep rise of both $B(M1)$ and $B(E2)$ values.
Although their ratios are still decreasing with increasing spin, a kink is present around $I=12\hbar$. 

\begin{table}[!htbp]
 \caption{Cranking frequencies $\hbar\omega$, angular momenta $I$, excitation energies $E$, and reduced transition probabilities for the positive-parity configuration $\pi g_{9/2}^{-1}\otimes\nu h_{11/2}^{2}(gd)^1$ as calculated by the 3DTAC method based on the covariant DFT. The excitation energies are renormalized to the ground state.
  \label{table1}}
 \centering
 \begin{tabular}{cccccc}\hline
 $\hbar\omega$  & $I$       & $E$   & $B(M1)$      & $B(E2)$    & $B(M1)/B(E2)$  \\
 ~[MeV]         & [$\hbar$] & [MeV] &  [$\mu_N^2$] & [0.1 $e^2b^2$] & [$\mu_N^2/(e^2 b^2)$]  \\
 \hline
 0.30 &  14.9  &  1.69 & 1.16 & 0.85 & 13.7   \\
 0.35 &  15.7  &  1.96 & 1.11 & 0.85 & 13.1   \\
 0.40 &  16.5  &  2.26 & 1.06 & 0.85 & 12.5   \\
 0.45 &  17.3  &  2.60 & 1.01 & 0.84 & 12.0   \\
 0.50 &  18.1  &  2.99 & 0.96 & 0.85 & 11.3   \\
 0.55 &  18.9  &  3.40 & 0.91 & 0.81 & 11.2   \\
 0.60 &  19.7  &  3.85 & 0.86 & 0.79 & 10.9   \\
 0.65 &  20.5  &  4.34 & 0.81 & 0.76 & 10.7  \\
 0.70 &  22.1  &  5.45 & 1.32 & 1.01 & 13.1   \\
\hline
 \end{tabular}
\end{table}

A second chiral band is predicted in $^{106}$Rh from the results of the present calculations, and it is based on the positive-parity configuration $\pi g_{9/2}^{-1}\otimes\nu h_{11/2}^{2}(gd)^1$.  
Note that this configuration is different from that with positive parity reported in Ref.~\cite{Meng2006Phys.Rev.C37303}, where the two neutrons in the $h_{11/2}$ shell are anti-aligned due to the imposed time-reversal symmetry. 
However, the breaking of this symmetry in the present cranking calculations enables the alignment of the two valence neutrons. 
This provides the angular momentum along the short axis, and plays a crucial role in the generation of the chiral band. 

The calculated angular momenta, excitation energies, and the reduced transition probabilities are listed in Table~\ref{table1}. 
Features, such as the kinks in the $\omega$-$I$ relation and the $B(M1)/B(E2)$ ratios, similar to those seen in the negative-parity band, are found for this band as well.  
Quantitatively, the excitation energy of this positive-parity band is roughly 1 MeV higher than that of the negative-parity one at any given rotational frequency. 
The angular momentum $I$ also becomes higher because of the alignment of the two $h_{11/2}$ neutrons, and its linear increase with frequency is marked by a steep rise of slope above $\hbar\omega=0.66$ MeV.
This is due to the abrupt change of the $\phi$ values (see Fig.~\ref{fig3}), and for the same reason, both $B(M1)$ and $B(E2)$ values increase rapidly above $\hbar\omega=0.66$ MeV. 

It is worthwhile to mention that similar calculations have also been carried out for the neighboring nucleus $^{105}$Rh, in which two chiral bands are observed based on the configurations $\pi g_{9/2}^{-1}\otimes\nu h_{11/2}^{1}(gd)^1$ and $\pi g_{9/2}^{-1}\otimes\nu h_{11/2}^{2}$, respectively~\cite{Alcantara-Nunez2004Phys.Rev.C24317,Timar2004Phys.Lett.B178}. 
It is found that the experimental data for both bands can be reproduced well by the present approach of 3DTAC-CDFT. 
For the relation between the angular momentum and rotational frequency, the present calculations provide a similar, or even better, agreement with the data in comparison with the results given by the previous shell correction tilted axis cranking method~\cite{Alcantara-Nunez2004Phys.Rev.C24317,Timar2004Phys.Lett.B178}.
In fact, several candidates of chiral bands have also been observed in other Rh isotopes. 
In particular, the phenomenon of multiple chirality based on the same configurations has been reported in $^{103}$Rh~\cite{Kuti2014Phys.Rev.Lett.32501}. 
Therefore, it would be interesting to perform a systematical investigation of nuclear chirality in Rh isotopes in the future. 
This will be discussed together with a detailed introduction of the formalism of 3DTAC-CDFT in a forthcoming paper. 

\section{\label{sec4} Summary}
In summary, covariant density functional theory and three-dimensional tilted axis cranking are used to investigate multiple chirality in nuclear rotation for the first time in a fully self-consistent and microscopic way. 
Two distinct sets of chiral solutions based on the respective configurations $\pi g_{9/2}^{-1}\otimes\nu h_{11/2}^{1}$ and $\pi g_{9/2}^{-1}\otimes\nu h_{11/2}^{2}(gd)^1$ have been uncovered in the nucleus $^{106}$Rh, which are the first examples for multiple chirality found by means of covariant DFT. 
A transition between planar and chiral rotation has been found for both configurations, while in the chiral regime, the tunneling between the left and the right-handed orientations could be substantial due to the soft Routhians. 
The calculated energy spectrum and $B(M1)/B(E2)$ ratios for the negative-parity band are in good agreement with the corresponding experimental data.
This demonstrates the predictive power of the present investigation and, thus, the potential for observation of the other predicted positive-parity chiral band appears to be good.

\section*{Acknowledgments}
The author is grateful to Q. B. Chen, R. V. F. Janssens, J. Meng, and S. Q. Zhang for helpful discussions and careful readings of the manuscript.
This work is supported by U.S. Department of Energy (DOE), Office of Science, Office of Nuclear Physics, under contract DE-AC02-06CH11357. 
Computational resources have been provided by the Laboratory Computing Resource Center at Argonne National Laboratory.

\section*{References}


\begin{thebibliography}{51}
\expandafter\ifx\csname natexlab\endcsname\relax\def\natexlab#1{#1}\fi
\providecommand{\url}[1]{\texttt{#1}}
\providecommand{\href}[2]{#2}
\providecommand{\path}[1]{#1}
\providecommand{\DOIprefix}{doi:}
\providecommand{\ArXivprefix}{arXiv:}
\providecommand{\URLprefix}{URL: }
\providecommand{\Pubmedprefix}{pmid:}
\providecommand{\doi}[1]{\href{http://dx.doi.org/#1}{\path{#1}}}
\providecommand{\Pubmed}[1]{\href{pmid:#1}{\path{#1}}}
\providecommand{\bibinfo}[2]{#2}
\ifx\xfnm\relax \def\xfnm[#1]{\unskip,\space#1}\fi
\bibitem[{Frauendorf and Meng(1997)}]{Frauendorf1997Nucl.Phys.A131}
\bibinfo{author}{S.~Frauendorf}, \bibinfo{author}{J.~Meng},
  \bibinfo{journal}{Nucl. Phys. A} \bibinfo{volume}{617} (\bibinfo{year}{1997})
  \bibinfo{pages}{131--147}. \DOIprefix\doi{10.1016/S0375-9474(97)00004-3}.
\bibitem[{Frauendorf(2001)}]{Frauendorf2001Rev.Mod.Phys.463}
\bibinfo{author}{S.~Frauendorf}, \bibinfo{journal}{Rev. Mod. Phys.}
  \bibinfo{volume}{73} (\bibinfo{year}{2001}) \bibinfo{pages}{463}.
  \DOIprefix\doi{10.1103/RevModPhys.73.463}.
\bibitem[{Meng and Zhang(2010)}]{Meng2010JPhysG.37.64025}
\bibinfo{author}{J.~Meng}, \bibinfo{author}{S.~Q. Zhang}, \bibinfo{journal}{J.
  Phys. G.} \bibinfo{volume}{37} (\bibinfo{year}{2010})
  \bibinfo{pages}{064025}.
\bibitem[{Starosta et~al.(2001)Starosta, Koike, Chiara, Fossan, LaFosse, Hecht,
  Beausang, Caprio, Cooper, Kr\"ucken, Novak, Zamfir, Zyromski, Hartley,
  Balabanski, Zhang, Frauendorf, and Dimitrov}]{Starosta2001Phys.Rev.Lett.971}
\bibinfo{author}{K.~Starosta}, \bibinfo{author}{T.~Koike},
  \bibinfo{author}{C.~J. Chiara}, \bibinfo{author}{D.~B. Fossan},
  \bibinfo{author}{D.~R. LaFosse}, \bibinfo{author}{A.~A. Hecht},
  \bibinfo{author}{C.~W. Beausang}, \bibinfo{author}{M.~A. Caprio},
  \bibinfo{author}{J.~R. Cooper}, \bibinfo{author}{R.~Kr\"ucken},
  \bibinfo{author}{J.~R. Novak}, \bibinfo{author}{N.~V. Zamfir},
  \bibinfo{author}{K.~E. Zyromski}, \bibinfo{author}{D.~J. Hartley},
  \bibinfo{author}{D.~L. Balabanski}, \bibinfo{author}{J.-y. Zhang},
  \bibinfo{author}{S.~Frauendorf}, \bibinfo{author}{V.~I. Dimitrov},
  \bibinfo{journal}{Phys. Rev. Lett.} \bibinfo{volume}{86}
  (\bibinfo{year}{2001}) \bibinfo{pages}{971--974}.
  \DOIprefix\doi{10.1103/PhysRevLett.86.971}.
\bibitem[{Zhu et~al.(2003)Zhu, Garg, Nayak, Ghugre, Pattabiraman, Fossan,
  Koike, Starosta, Vaman, Janssens, Chakrawarthy, Whitehead, Macchiavelli, and
  Frauendorf}]{Zhu2003Phys.Rev.Lett.132501}
\bibinfo{author}{S.~Zhu}, \bibinfo{author}{U.~Garg}, \bibinfo{author}{B.~K.
  Nayak}, \bibinfo{author}{S.~S. Ghugre}, \bibinfo{author}{N.~S. Pattabiraman},
  \bibinfo{author}{D.~B. Fossan}, \bibinfo{author}{T.~Koike},
  \bibinfo{author}{K.~Starosta}, \bibinfo{author}{C.~Vaman},
  \bibinfo{author}{R.~V.~F. Janssens}, \bibinfo{author}{R.~S. Chakrawarthy},
  \bibinfo{author}{M.~Whitehead}, \bibinfo{author}{A.~O. Macchiavelli},
  \bibinfo{author}{S.~Frauendorf}, \bibinfo{journal}{Phys. Rev. Lett.}
  \bibinfo{volume}{91} (\bibinfo{year}{2003}) \bibinfo{pages}{132501}.
  \DOIprefix\doi{10.1103/PhysRevLett.91.132501}.
\bibitem[{Vaman et~al.(2004)Vaman, Fossan, Koike, Starosta, Lee, and
  Macchiavelli}]{Vaman2004Phys.Rev.Lett.32501}
\bibinfo{author}{C.~Vaman}, \bibinfo{author}{D.~B. Fossan},
  \bibinfo{author}{T.~Koike}, \bibinfo{author}{K.~Starosta},
  \bibinfo{author}{I.~Y. Lee}, \bibinfo{author}{A.~O. Macchiavelli},
  \bibinfo{journal}{Phys. Rev. Lett.} \bibinfo{volume}{92}
  (\bibinfo{year}{2004}) \bibinfo{pages}{032501}.
  \DOIprefix\doi{10.1103/PhysRevLett.92.032501}.
\bibitem[{Joshi et~al.(2004)Joshi, Jenkins, Raddon, Simons, Wadsworth,
  Wilkinson, Fossan, Koike, Starosta, Vaman, Timár, Dombrádi, Krasznahorkay,
  Molnár, Sohler, Zolnai, Algora, Paul, Rainovski, Gizon, Gizon, Bednarczyk,
  Curien, Duchêne, and Scheurer}]{Joshi2004Phys.Lett.B135}
\bibinfo{author}{P.~Joshi}, \bibinfo{author}{D.~Jenkins},
  \bibinfo{author}{P.~Raddon}, \bibinfo{author}{A.~Simons},
  \bibinfo{author}{R.~Wadsworth}, \bibinfo{author}{A.~Wilkinson},
  \bibinfo{author}{D.~Fossan}, \bibinfo{author}{T.~Koike},
  \bibinfo{author}{K.~Starosta}, \bibinfo{author}{C.~Vaman},
  \bibinfo{author}{J.~Timár}, \bibinfo{author}{Z.~Dombrádi},
  \bibinfo{author}{A.~Krasznahorkay}, \bibinfo{author}{J.~Molnár},
  \bibinfo{author}{D.~Sohler}, \bibinfo{author}{L.~Zolnai},
  \bibinfo{author}{A.~Algora}, \bibinfo{author}{E.~Paul},
  \bibinfo{author}{G.~Rainovski}, \bibinfo{author}{A.~Gizon},
  \bibinfo{author}{J.~Gizon}, \bibinfo{author}{P.~Bednarczyk},
  \bibinfo{author}{D.~Curien}, \bibinfo{author}{G.~Duchêne},
  \bibinfo{author}{J.~Scheurer}, \bibinfo{journal}{Phys. Lett. B}
  \bibinfo{volume}{595} (\bibinfo{year}{2004}) \bibinfo{pages}{135--142}.
  \DOIprefix\doi{10.1016/j.physletb.2004.05.066}.
\bibitem[{Alc\'antara-N\'u\~nez et~al.(2004)Alc\'antara-N\'u\~nez, Oliveira,
  Cybulska, Medina, Rao, Ribas, Rizzutto, Seale, Falla-Sotelo, Wiedemann,
  Dimitrov, and Frauendorf}]{Alcantara-Nunez2004Phys.Rev.C24317}
\bibinfo{author}{J.~A. Alc\'antara-N\'u\~nez}, \bibinfo{author}{J.~R.~B.
  Oliveira}, \bibinfo{author}{E.~W. Cybulska}, \bibinfo{author}{N.~H. Medina},
  \bibinfo{author}{M.~N. Rao}, \bibinfo{author}{R.~V. Ribas},
  \bibinfo{author}{M.~A. Rizzutto}, \bibinfo{author}{W.~A. Seale},
  \bibinfo{author}{F.~Falla-Sotelo}, \bibinfo{author}{K.~T. Wiedemann},
  \bibinfo{author}{V.~I. Dimitrov}, \bibinfo{author}{S.~Frauendorf},
  \bibinfo{journal}{Phys. Rev. C} \bibinfo{volume}{69} (\bibinfo{year}{2004})
  \bibinfo{pages}{024317}. \DOIprefix\doi{10.1103/PhysRevC.69.024317}.
\bibitem[{Tim{\'{a}}r et~al.(2004)Tim{\'{a}}r, Joshi, Starosta, Dimitrov,
  Fossan, Moln{\'{a}}r, Sohler, Wadsworth, Algora, Bednarczyk, Curien,
  Dombr{\'{a}}di, Duchene, Gizon, Gizon, Jenkins, Koike, Krasznahorkay, Paul,
  Raddon, Rainovski, Scheurer, Simons, Vaman, Wilkinson, Zolnai, and
  Frauendorf}]{Timar2004Phys.Lett.B178}
\bibinfo{author}{J.~Tim{\'{a}}r}, \bibinfo{author}{P.~Joshi},
  \bibinfo{author}{K.~Starosta}, \bibinfo{author}{V.~Dimitrov},
  \bibinfo{author}{D.~Fossan}, \bibinfo{author}{J.~Moln{\'{a}}r},
  \bibinfo{author}{D.~Sohler}, \bibinfo{author}{R.~Wadsworth},
  \bibinfo{author}{A.~Algora}, \bibinfo{author}{P.~Bednarczyk},
  \bibinfo{author}{D.~Curien}, \bibinfo{author}{Z.~Dombr{\'{a}}di},
  \bibinfo{author}{G.~Duchene}, \bibinfo{author}{A.~Gizon},
  \bibinfo{author}{J.~Gizon}, \bibinfo{author}{D.~Jenkins},
  \bibinfo{author}{T.~Koike}, \bibinfo{author}{A.~Krasznahorkay},
  \bibinfo{author}{E.~Paul}, \bibinfo{author}{P.~Raddon},
  \bibinfo{author}{G.~Rainovski}, \bibinfo{author}{J.~Scheurer},
  \bibinfo{author}{A.~Simons}, \bibinfo{author}{C.~Vaman},
  \bibinfo{author}{A.~Wilkinson}, \bibinfo{author}{L.~Zolnai},
  \bibinfo{author}{S.~Frauendorf}, \bibinfo{journal}{Phys. Lett. B}
  \bibinfo{volume}{598} (\bibinfo{year}{2004}) \bibinfo{pages}{178--187}.
  \DOIprefix\doi{10.1016/j.physletb.2004.07.050}.
\bibitem[{Grodner et~al.(2006)Grodner, Srebrny, Pasternak, Zalewska, Morek,
  Droste, Mierzejewski, Kowalczyk, Kownacki, Kisieli\ifmmode~\acute{n}\else
  \'{n}\fi{}ski, Rohozi\ifmmode~\acute{n}\else \'{n}\fi{}ski, Koike, Starosta,
  Kordyasz, Napiorkowski, Woli\ifmmode \acute{n}\else~\'{n}\fi{}ska Cichocka,
  Ruchowska, P\l{}\'ociennik, and Perkowski}]{Grodner2006Phys.Rev.Lett.172501}
\bibinfo{author}{E.~Grodner}, \bibinfo{author}{J.~Srebrny},
  \bibinfo{author}{A.~A. Pasternak}, \bibinfo{author}{I.~Zalewska},
  \bibinfo{author}{T.~Morek}, \bibinfo{author}{C.~Droste},
  \bibinfo{author}{J.~Mierzejewski}, \bibinfo{author}{M.~Kowalczyk},
  \bibinfo{author}{J.~Kownacki},
  \bibinfo{author}{M.~Kisieli\ifmmode~\acute{n}\else \'{n}\fi{}ski},
  \bibinfo{author}{S.~G. Rohozi\ifmmode~\acute{n}\else \'{n}\fi{}ski},
  \bibinfo{author}{T.~Koike}, \bibinfo{author}{K.~Starosta},
  \bibinfo{author}{A.~Kordyasz}, \bibinfo{author}{P.~J. Napiorkowski},
  \bibinfo{author}{M.~Woli\ifmmode \acute{n}\else~\'{n}\fi{}ska Cichocka},
  \bibinfo{author}{E.~Ruchowska}, \bibinfo{author}{W.~P\l{}\'ociennik},
  \bibinfo{author}{J.~Perkowski}, \bibinfo{journal}{Phys. Rev. Lett.}
  \bibinfo{volume}{97} (\bibinfo{year}{2006}) \bibinfo{pages}{172501}.
  \DOIprefix\doi{10.1103/PhysRevLett.97.172501}.
\bibitem[{Joshi et~al.(2007)Joshi, Carpenter, Fossan, Koike, Paul, Rainovski,
  Starosta, Vaman, and Wadsworth}]{Joshi2007Phys.Rev.Lett.102501}
\bibinfo{author}{P.~Joshi}, \bibinfo{author}{M.~P. Carpenter},
  \bibinfo{author}{D.~B. Fossan}, \bibinfo{author}{T.~Koike},
  \bibinfo{author}{E.~S. Paul}, \bibinfo{author}{G.~Rainovski},
  \bibinfo{author}{K.~Starosta}, \bibinfo{author}{C.~Vaman},
  \bibinfo{author}{R.~Wadsworth}, \bibinfo{journal}{Phys. Rev. Lett.}
  \bibinfo{volume}{98} (\bibinfo{year}{2007}) \bibinfo{pages}{102501}.
  \DOIprefix\doi{10.1103/PhysRevLett.98.102501}.
\bibitem[{Mukhopadhyay et~al.(2007)Mukhopadhyay, Almehed, Garg, Frauendorf, Li,
  Rao, Wang, Ghugre, Carpenter, Gros, Hecht, Janssens, Kondev, Lauritsen,
  Seweryniak, and Zhu}]{Mukhopadhyay2007Phys.Rev.Lett.172501}
\bibinfo{author}{S.~Mukhopadhyay}, \bibinfo{author}{D.~Almehed},
  \bibinfo{author}{U.~Garg}, \bibinfo{author}{S.~Frauendorf},
  \bibinfo{author}{T.~Li}, \bibinfo{author}{P.~V.~M. Rao},
  \bibinfo{author}{X.~Wang}, \bibinfo{author}{S.~S. Ghugre},
  \bibinfo{author}{M.~P. Carpenter}, \bibinfo{author}{S.~Gros},
  \bibinfo{author}{A.~Hecht}, \bibinfo{author}{R.~V.~F. Janssens},
  \bibinfo{author}{F.~G. Kondev}, \bibinfo{author}{T.~Lauritsen},
  \bibinfo{author}{D.~Seweryniak}, \bibinfo{author}{S.~Zhu},
  \bibinfo{journal}{Phys. Rev. Lett.} \bibinfo{volume}{99}
  (\bibinfo{year}{2007}) \bibinfo{pages}{172501}.
  \DOIprefix\doi{10.1103/PhysRevLett.99.172501}.
\bibitem[{Grodner et~al.(2011)Grodner, Sankowska, Morek, Rohozi{\'{n}}ski,
  Droste, Srebrny, Pasternak, Kisieli{\'{n}}ski, Kowalczyk, Kownacki,
  Mierzejewski, Kr{\'{o}}l, and Wrzosek}]{Grodner2011Phys.Lett.B46}
\bibinfo{author}{E.~Grodner}, \bibinfo{author}{I.~Sankowska},
  \bibinfo{author}{T.~Morek}, \bibinfo{author}{S.~Rohozi{\'{n}}ski},
  \bibinfo{author}{C.~Droste}, \bibinfo{author}{J.~Srebrny},
  \bibinfo{author}{A.~Pasternak}, \bibinfo{author}{M.~Kisieli{\'{n}}ski},
  \bibinfo{author}{M.~Kowalczyk}, \bibinfo{author}{J.~Kownacki},
  \bibinfo{author}{J.~Mierzejewski}, \bibinfo{author}{A.~Kr{\'{o}}l},
  \bibinfo{author}{K.~Wrzosek}, \bibinfo{journal}{Phys. Lett. B}
  \bibinfo{volume}{703} (\bibinfo{year}{2011}) \bibinfo{pages}{46--50}.
  \DOIprefix\doi{10.1016/j.physletb.2011.07.062}.
\bibitem[{Wang et~al.(2011)Wang, Qi, Liu, Zhang, Hua, Li, Chen, Zhu, Meng,
  Wyngaardt, Papka, Ibrahim, Bark, Datta, Lawrie, Lawrie, Majola, Masiteng,
  Mullins, G{\'{a}}l, Kalinka, Moln{\'{a}}r, Nyak{\'{o}}, Tim{\'{a}}r,
  Juh{\'{a}}sz, and Schwengner}]{Wang2011Phys.Lett.B40}
\bibinfo{author}{S.~Wang}, \bibinfo{author}{B.~Qi}, \bibinfo{author}{L.~Liu},
  \bibinfo{author}{S.~Zhang}, \bibinfo{author}{H.~Hua},
  \bibinfo{author}{X.~Li}, \bibinfo{author}{Y.~Chen}, \bibinfo{author}{L.~Zhu},
  \bibinfo{author}{J.~Meng}, \bibinfo{author}{S.~Wyngaardt},
  \bibinfo{author}{P.~Papka}, \bibinfo{author}{T.~Ibrahim},
  \bibinfo{author}{R.~Bark}, \bibinfo{author}{P.~Datta},
  \bibinfo{author}{E.~Lawrie}, \bibinfo{author}{J.~Lawrie},
  \bibinfo{author}{S.~Majola}, \bibinfo{author}{P.~Masiteng},
  \bibinfo{author}{S.~Mullins}, \bibinfo{author}{J.~G{\'{a}}l},
  \bibinfo{author}{G.~Kalinka}, \bibinfo{author}{J.~Moln{\'{a}}r},
  \bibinfo{author}{B.~Nyak{\'{o}}}, \bibinfo{author}{J.~Tim{\'{a}}r},
  \bibinfo{author}{K.~Juh{\'{a}}sz}, \bibinfo{author}{R.~Schwengner},
  \bibinfo{journal}{Phys. Lett. B} \bibinfo{volume}{703} (\bibinfo{year}{2011})
  \bibinfo{pages}{40--45}. \DOIprefix\doi{10.1016/j.physletb.2011.07.055}.
\bibitem[{Ayangeakaa et~al.(2013)Ayangeakaa, Garg, Anthony, Frauendorf, Matta,
  Nayak, Patel, Chen, Zhang, Zhao, Qi, Meng, Janssens, Carpenter, Chiara,
  Kondev, Lauritsen, Seweryniak, Zhu, Ghugre, and
  Palit}]{Ayangeakaa2013Phys.Rev.Lett.172504}
\bibinfo{author}{A.~D. Ayangeakaa}, \bibinfo{author}{U.~Garg},
  \bibinfo{author}{M.~D. Anthony}, \bibinfo{author}{S.~Frauendorf},
  \bibinfo{author}{J.~T. Matta}, \bibinfo{author}{B.~K. Nayak},
  \bibinfo{author}{D.~Patel}, \bibinfo{author}{Q.~B. Chen},
  \bibinfo{author}{S.~Q. Zhang}, \bibinfo{author}{P.~W. Zhao},
  \bibinfo{author}{B.~Qi}, \bibinfo{author}{J.~Meng}, \bibinfo{author}{R.~V.~F.
  Janssens}, \bibinfo{author}{M.~P. Carpenter}, \bibinfo{author}{C.~J. Chiara},
  \bibinfo{author}{F.~G. Kondev}, \bibinfo{author}{T.~Lauritsen},
  \bibinfo{author}{D.~Seweryniak}, \bibinfo{author}{S.~Zhu},
  \bibinfo{author}{S.~S. Ghugre}, \bibinfo{author}{R.~Palit},
  \bibinfo{journal}{Phys. Rev. Lett.} \bibinfo{volume}{110}
  (\bibinfo{year}{2013}) \bibinfo{pages}{172504}.
  \DOIprefix\doi{10.1103/PhysRevLett.110.172504}.
\bibitem[{Kuti et~al.(2014)Kuti, Chen, Tim\'ar, Sohler, Zhang, Zhang, Zhao,
  Meng, Starosta, Koike, Paul, Fossan, and Vaman}]{Kuti2014Phys.Rev.Lett.32501}
\bibinfo{author}{I.~Kuti}, \bibinfo{author}{Q.~B. Chen},
  \bibinfo{author}{J.~Tim\'ar}, \bibinfo{author}{D.~Sohler},
  \bibinfo{author}{S.~Q. Zhang}, \bibinfo{author}{Z.~H. Zhang},
  \bibinfo{author}{P.~W. Zhao}, \bibinfo{author}{J.~Meng},
  \bibinfo{author}{K.~Starosta}, \bibinfo{author}{T.~Koike},
  \bibinfo{author}{E.~S. Paul}, \bibinfo{author}{D.~B. Fossan},
  \bibinfo{author}{C.~Vaman}, \bibinfo{journal}{Phys. Rev. Lett.}
  \bibinfo{volume}{113} (\bibinfo{year}{2014}) \bibinfo{pages}{032501}.
  \DOIprefix\doi{10.1103/PhysRevLett.113.032501}.
\bibitem[{Tonev et~al.(2014)Tonev, Yavahchova, Goutev, de~Angelis, Petkov,
  Bhowmik, Singh, Muralithar, Madhavan, Kumar, Kumar~Raju, Kaur, Mohanto,
  Singh, Kaur, Garg, Shukla, Marinov, and Brant}]{Tonev2014Phys.Rev.Lett.52501}
\bibinfo{author}{D.~Tonev}, \bibinfo{author}{M.~S. Yavahchova},
  \bibinfo{author}{N.~Goutev}, \bibinfo{author}{G.~de~Angelis},
  \bibinfo{author}{P.~Petkov}, \bibinfo{author}{R.~K. Bhowmik},
  \bibinfo{author}{R.~P. Singh}, \bibinfo{author}{S.~Muralithar},
  \bibinfo{author}{N.~Madhavan}, \bibinfo{author}{R.~Kumar},
  \bibinfo{author}{M.~Kumar~Raju}, \bibinfo{author}{J.~Kaur},
  \bibinfo{author}{G.~Mohanto}, \bibinfo{author}{A.~Singh},
  \bibinfo{author}{N.~Kaur}, \bibinfo{author}{R.~Garg},
  \bibinfo{author}{A.~Shukla}, \bibinfo{author}{T.~K. Marinov},
  \bibinfo{author}{S.~Brant}, \bibinfo{journal}{Phys. Rev. Lett.}
  \bibinfo{volume}{112} (\bibinfo{year}{2014}) \bibinfo{pages}{052501}.
  \DOIprefix\doi{10.1103/PhysRevLett.112.052501}.
\bibitem[{Liu et~al.(2016)Liu, Wang, Bark, Zhang, Meng, Qi, Jones, Wyngaardt,
  Zhao, Xu, Zhou, Wang, Sun, Liu, Li, Zhang, Jia, Li, Hua, Chen, Xiao, Li, Zhu,
  Bucher, Dinoko, Easton, Juh\'asz, Kamblawe, Khaleel, Khumalo, Lawrie, Lawrie,
  Majola, Mullins, Murray, Ndayishimye, Negi, Noncolela, Ntshangase, Nyak\'o,
  Orce, Papka, Sharpey-Schafer, Shirinda, Sithole, Stankiewicz, and
  Wiedeking}]{Liu2016Phys.Rev.Lett.112501}
\bibinfo{author}{C.~Liu}, \bibinfo{author}{S.~Y. Wang}, \bibinfo{author}{R.~A.
  Bark}, \bibinfo{author}{S.~Q. Zhang}, \bibinfo{author}{J.~Meng},
  \bibinfo{author}{B.~Qi}, \bibinfo{author}{P.~Jones}, \bibinfo{author}{S.~M.
  Wyngaardt}, \bibinfo{author}{J.~Zhao}, \bibinfo{author}{C.~Xu},
  \bibinfo{author}{S.-G. Zhou}, \bibinfo{author}{S.~Wang},
  \bibinfo{author}{D.~P. Sun}, \bibinfo{author}{L.~Liu}, \bibinfo{author}{Z.~Q.
  Li}, \bibinfo{author}{N.~B. Zhang}, \bibinfo{author}{H.~Jia},
  \bibinfo{author}{X.~Q. Li}, \bibinfo{author}{H.~Hua}, \bibinfo{author}{Q.~B.
  Chen}, \bibinfo{author}{Z.~G. Xiao}, \bibinfo{author}{H.~J. Li},
  \bibinfo{author}{L.~H. Zhu}, \bibinfo{author}{T.~D. Bucher},
  \bibinfo{author}{T.~Dinoko}, \bibinfo{author}{J.~Easton},
  \bibinfo{author}{K.~Juh\'asz}, \bibinfo{author}{A.~Kamblawe},
  \bibinfo{author}{E.~Khaleel}, \bibinfo{author}{N.~Khumalo},
  \bibinfo{author}{E.~A. Lawrie}, \bibinfo{author}{J.~J. Lawrie},
  \bibinfo{author}{S.~N.~T. Majola}, \bibinfo{author}{S.~M. Mullins},
  \bibinfo{author}{S.~Murray}, \bibinfo{author}{J.~Ndayishimye},
  \bibinfo{author}{D.~Negi}, \bibinfo{author}{S.~P. Noncolela},
  \bibinfo{author}{S.~S. Ntshangase}, \bibinfo{author}{B.~M. Nyak\'o},
  \bibinfo{author}{J.~N. Orce}, \bibinfo{author}{P.~Papka},
  \bibinfo{author}{J.~F. Sharpey-Schafer}, \bibinfo{author}{O.~Shirinda},
  \bibinfo{author}{P.~Sithole}, \bibinfo{author}{M.~A. Stankiewicz},
  \bibinfo{author}{M.~Wiedeking}, \bibinfo{journal}{Phys. Rev. Lett.}
  \bibinfo{volume}{116} (\bibinfo{year}{2016}) \bibinfo{pages}{112501}.
  \DOIprefix\doi{10.1103/PhysRevLett.116.112501}.
\bibitem[{Meng et~al.(2006)Meng, Peng, Zhang, and
  Zhou}]{Meng2006Phys.Rev.C37303}
\bibinfo{author}{J.~Meng}, \bibinfo{author}{J.~Peng}, \bibinfo{author}{S.~Q.
  Zhang}, \bibinfo{author}{S.-G. Zhou}, \bibinfo{journal}{Phys. Rev. C}
  \bibinfo{volume}{73} (\bibinfo{year}{2006}) \bibinfo{pages}{037303}.
  \DOIprefix\doi{10.1103/PhysRevC.73.037303}.
\bibitem[{Peng et~al.(2008)Peng, Sagawa, Zhang, Yao, Zhang, and
  Meng}]{Peng2008Phys.Rev.C24309}
\bibinfo{author}{J.~Peng}, \bibinfo{author}{H.~Sagawa}, \bibinfo{author}{S.~Q.
  Zhang}, \bibinfo{author}{J.~M. Yao}, \bibinfo{author}{Y.~Zhang},
  \bibinfo{author}{J.~Meng}, \bibinfo{journal}{Phys. Rev. C}
  \bibinfo{volume}{77} (\bibinfo{year}{2008}) \bibinfo{pages}{024309}.
  \DOIprefix\doi{10.1103/PhysRevC.77.024309}.
\bibitem[{Yao et~al.(2009)Yao, Qi, Zhang, Peng, Wang, and
  Meng}]{Yao2009Phys.Rev.C67302}
\bibinfo{author}{J.~M. Yao}, \bibinfo{author}{B.~Qi}, \bibinfo{author}{S.~Q.
  Zhang}, \bibinfo{author}{J.~Peng}, \bibinfo{author}{S.~Y. Wang},
  \bibinfo{author}{J.~Meng}, \bibinfo{journal}{Phys. Rev. C}
  \bibinfo{volume}{79} (\bibinfo{year}{2009}) \bibinfo{pages}{067302}.
  \DOIprefix\doi{10.1103/PhysRevC.79.067302}.
\bibitem[{Li et~al.(2011)Li, Zhang, and Meng}]{Li2011Phys.Rev.C37301}
\bibinfo{author}{J.~Li}, \bibinfo{author}{S.~Q. Zhang},
  \bibinfo{author}{J.~Meng}, \bibinfo{journal}{Phys. Rev. C}
  \bibinfo{volume}{83} (\bibinfo{year}{2011}) \bibinfo{pages}{037301}.
  \DOIprefix\doi{10.1103/PhysRevC.83.037301}.
\bibitem[{Peng et~al.(2003)Peng, Meng, and Zhang}]{Peng2003Phys.Rev.C44324}
\bibinfo{author}{J.~Peng}, \bibinfo{author}{J.~Meng}, \bibinfo{author}{S.~Q.
  Zhang}, \bibinfo{journal}{Phys. Rev. C} \bibinfo{volume}{68}
  (\bibinfo{year}{2003}) \bibinfo{pages}{044324}.
  \DOIprefix\doi{10.1103/PhysRevC.68.044324}.
\bibitem[{Koike et~al.(2004)Koike, Starosta, and
  Hamamoto}]{Koike2004Phys.Rev.Lett.172502}
\bibinfo{author}{T.~Koike}, \bibinfo{author}{K.~Starosta},
  \bibinfo{author}{I.~Hamamoto}, \bibinfo{journal}{Phys. Rev. Lett.}
  \bibinfo{volume}{93} (\bibinfo{year}{2004}) \bibinfo{pages}{172502}.
  \DOIprefix\doi{10.1103/PhysRevLett.93.172502}.
\bibitem[{Zhang et~al.(2007)Zhang, Qi, Wang, and
  Meng}]{Zhang2007Phys.Rev.C44307}
\bibinfo{author}{S.~Q. Zhang}, \bibinfo{author}{B.~Qi}, \bibinfo{author}{S.~Y.
  Wang}, \bibinfo{author}{J.~Meng}, \bibinfo{journal}{Phys. Rev. C}
  \bibinfo{volume}{75} (\bibinfo{year}{2007}) \bibinfo{pages}{044307}.
  \DOIprefix\doi{10.1103/PhysRevC.75.044307}.
\bibitem[{Qi et~al.(2009)Qi, Zhang, Meng, Wang, and
  Frauendorf}]{Qi2009Phys.Lett.B175}
\bibinfo{author}{B.~Qi}, \bibinfo{author}{S.~Zhang}, \bibinfo{author}{J.~Meng},
  \bibinfo{author}{S.~Wang}, \bibinfo{author}{S.~Frauendorf},
  \bibinfo{journal}{Phys. Lett. B} \bibinfo{volume}{675} (\bibinfo{year}{2009})
  \bibinfo{pages}{175--180}. \DOIprefix\doi{10.1016/j.physletb.2009.02.061}.
\bibitem[{Dimitrov et~al.(2000)Dimitrov, Frauendorf, and
  D\"onau}]{Dimitrov2000Phys.Rev.Lett.5732}
\bibinfo{author}{V.~I. Dimitrov}, \bibinfo{author}{S.~Frauendorf},
  \bibinfo{author}{F.~D\"onau}, \bibinfo{journal}{Phys. Rev. Lett.}
  \bibinfo{volume}{84} (\bibinfo{year}{2000}) \bibinfo{pages}{5732--5735}.
  \DOIprefix\doi{10.1103/PhysRevLett.84.5732}.
\bibitem[{Olbratowski et~al.(2004)Olbratowski, Dobaczewski, Dudek, and
  P\l{}\'ociennik}]{Olbratowski2004Phys.Rev.Lett.52501}
\bibinfo{author}{P.~Olbratowski}, \bibinfo{author}{J.~Dobaczewski},
  \bibinfo{author}{J.~Dudek}, \bibinfo{author}{W.~P\l{}\'ociennik},
  \bibinfo{journal}{Phys. Rev. Lett.} \bibinfo{volume}{93}
  (\bibinfo{year}{2004}) \bibinfo{pages}{052501}.
  \DOIprefix\doi{10.1103/PhysRevLett.93.052501}.
\bibitem[{Lalazissis et~al.(2004)Lalazissis, Ring, and
  Vretenar}]{Lalazissis2004}
\bibinfo{editor}{G.~A. Lalazissis}, \bibinfo{editor}{P.~Ring},
  \bibinfo{editor}{D.~Vretenar} (Eds.), \bibinfo{title}{Extended Density
  Functionals in Nuclear Structure Physics}, \bibinfo{publisher}{Springer
  Berlin Heidelberg}, \bibinfo{year}{2004}. \DOIprefix\doi{10.1007/b95720}.
\bibitem[{Cohen et~al.(1992)Cohen, Furnstahl, and
  Griegel}]{Cohen1992Phys.Rev.C1881}
\bibinfo{author}{T.~D. Cohen}, \bibinfo{author}{R.~J. Furnstahl},
  \bibinfo{author}{D.~K. Griegel}, \bibinfo{journal}{Phys. Rev. C}
  \bibinfo{volume}{45} (\bibinfo{year}{1992}) \bibinfo{pages}{1881--1893}.
  \DOIprefix\doi{10.1103/PhysRevC.45.1881}.
\bibitem[{Koepf and Ring(1989)}]{Koepf1989Nucl.Phys.A61}
\bibinfo{author}{W.~Koepf}, \bibinfo{author}{P.~Ring}, \bibinfo{journal}{Nucl.
  Phys. A} \bibinfo{volume}{493} (\bibinfo{year}{1989})
  \bibinfo{pages}{61--82}. \DOIprefix\doi{10.1016/0375-9474(89)90532-0}.
\bibitem[{Afanasjev et~al.(1999)Afanasjev, Fossan, Lane, and
  Ragnarsson}]{Afanasjev1999Phys.Rep.1}
\bibinfo{author}{A.~V. Afanasjev}, \bibinfo{author}{D.~B. Fossan},
  \bibinfo{author}{G.~J. Lane}, \bibinfo{author}{I.~Ragnarsson},
  \bibinfo{journal}{Phys. Rep.} \bibinfo{volume}{322} (\bibinfo{year}{1999})
  \bibinfo{pages}{1--124}. \DOIprefix\doi{10.1016/S0370-1573(99)00035-6}.
\bibitem[{Meng et~al.(2013)Meng, Peng, Zhang, and Zhao}]{Meng2013Front.Phys.55}
\bibinfo{author}{J.~Meng}, \bibinfo{author}{J.~Peng}, \bibinfo{author}{S.-Q.
  Zhang}, \bibinfo{author}{P.-W. Zhao}, \bibinfo{journal}{Front. Phys.}
  \bibinfo{volume}{8} (\bibinfo{year}{2013}) \bibinfo{pages}{55--79}.
  \DOIprefix\doi{10.1007/s11467-013-0287-y}.
\bibitem[{Meng and Zhao(2016)}]{Meng2016PhysicaScripta53008}
\bibinfo{author}{J.~Meng}, \bibinfo{author}{P.~Zhao}, \bibinfo{journal}{Physica
  Scripta} \bibinfo{volume}{91} (\bibinfo{year}{2016}) \bibinfo{pages}{053008}.
  \DOIprefix\doi{10.1088/0031-8949/91/5/053008}.
\bibitem[{Zhao et~al.(2011)Zhao, Zhang, Peng, Liang, Ring, and
  Meng}]{Zhao2011Phys.Lett.B181}
\bibinfo{author}{P.~W. Zhao}, \bibinfo{author}{S.~Q. Zhang},
  \bibinfo{author}{J.~Peng}, \bibinfo{author}{H.~Z. Liang},
  \bibinfo{author}{P.~Ring}, \bibinfo{author}{J.~Meng}, \bibinfo{journal}{Phys.
  Lett. B} \bibinfo{volume}{699} (\bibinfo{year}{2011})
  \bibinfo{pages}{181--186}. \DOIprefix\doi{10.1016/j.physletb.2011.03.068}.
\bibitem[{Yu et~al.(2012)Yu, Zhao, Zhang, Ring, and
  Meng}]{Yu2012Phys.Rev.C24318}
\bibinfo{author}{L.~F. Yu}, \bibinfo{author}{P.~W. Zhao},
  \bibinfo{author}{S.~Q. Zhang}, \bibinfo{author}{P.~Ring},
  \bibinfo{author}{J.~Meng}, \bibinfo{journal}{Phys. Rev. C}
  \bibinfo{volume}{85} (\bibinfo{year}{2012}) \bibinfo{pages}{024318}.
  \DOIprefix\doi{10.1103/PhysRevC.85.024318}.
\bibitem[{Zhao et~al.(2011)Zhao, Peng, Liang, Ring, and
  Meng}]{Zhao2011Phys.Rev.Lett.122501}
\bibinfo{author}{P.~W. Zhao}, \bibinfo{author}{J.~Peng}, \bibinfo{author}{H.~Z.
  Liang}, \bibinfo{author}{P.~Ring}, \bibinfo{author}{J.~Meng},
  \bibinfo{journal}{Phys. Rev. Lett.} \bibinfo{volume}{107}
  (\bibinfo{year}{2011}) \bibinfo{pages}{122501}.
  \DOIprefix\doi{10.1103/PhysRevLett.107.122501}.
\bibitem[{Zhao et~al.(2012)Zhao, Peng, Liang, Ring, and
  Meng}]{Zhao2012Phys.Rev.C54310}
\bibinfo{author}{P.~W. Zhao}, \bibinfo{author}{J.~Peng}, \bibinfo{author}{H.~Z.
  Liang}, \bibinfo{author}{P.~Ring}, \bibinfo{author}{J.~Meng},
  \bibinfo{journal}{Phys. Rev. C} \bibinfo{volume}{85} (\bibinfo{year}{2012})
  \bibinfo{pages}{054310}. \DOIprefix\doi{10.1103/PhysRevC.85.054310}.
\bibitem[{Zhao et~al.(2015{\natexlab{a}})Zhao, Zhang, and
  Meng}]{Zhao2015Phys.Rev.C34319}
\bibinfo{author}{P.~W. Zhao}, \bibinfo{author}{S.~Q. Zhang},
  \bibinfo{author}{J.~Meng}, \bibinfo{journal}{Phys. Rev. C}
  \bibinfo{volume}{92} (\bibinfo{year}{2015}{\natexlab{a}})
  \bibinfo{pages}{034319}. \DOIprefix\doi{10.1103/PhysRevC.92.034319}.
\bibitem[{Zhao et~al.(2015{\natexlab{b}})Zhao, Itagaki, and
  Meng}]{Zhao2015Phys.Rev.Lett.22501}
\bibinfo{author}{P.~W. Zhao}, \bibinfo{author}{N.~Itagaki},
  \bibinfo{author}{J.~Meng}, \bibinfo{journal}{Phys. Rev. Lett.}
  \bibinfo{volume}{115} (\bibinfo{year}{2015}{\natexlab{b}})
  \bibinfo{pages}{022501}. \DOIprefix\doi{10.1103/PhysRevLett.115.022501}.
\bibitem[{Ring(1996)}]{Ring1996Prog.Part.Nucl.Phys.193}
\bibinfo{author}{P.~Ring}, \bibinfo{journal}{Prog. Part. Nucl. Phys.}
  \bibinfo{volume}{37} (\bibinfo{year}{1996}) \bibinfo{pages}{193--263}.
  \DOIprefix\doi{10.1016/0146-6410(96)00054-3}.
\bibitem[{Vretenar et~al.(2005)Vretenar, Afanasjev, Lalazissis, and
  Ring}]{Vretenar2005Phys.Rep.101}
\bibinfo{author}{D.~Vretenar}, \bibinfo{author}{A.~V. Afanasjev},
  \bibinfo{author}{G.~A. Lalazissis}, \bibinfo{author}{P.~Ring},
  \bibinfo{journal}{Phys. Rep.} \bibinfo{volume}{409} (\bibinfo{year}{2005})
  \bibinfo{pages}{101--259}. \DOIprefix\doi{10.1016/j.physrep.2004.10.001}.
\bibitem[{Meng et~al.(2006)Meng, Toki, Zhou, Zhang, Long, and
  Geng}]{Meng2006Prog.Part.Nucl.Phys.470}
\bibinfo{author}{J.~Meng}, \bibinfo{author}{H.~Toki},
  \bibinfo{author}{S.~Zhou}, \bibinfo{author}{S.~Zhang},
  \bibinfo{author}{W.~Long}, \bibinfo{author}{L.~Geng}, \bibinfo{journal}{Prog.
  Part. Nucl. Phys.} \bibinfo{volume}{57} (\bibinfo{year}{2006})
  \bibinfo{pages}{470--563}. \DOIprefix\doi{10.1016/j.ppnp.2005.06.001}.
\bibitem[{Nik\v{s}i\'{c} et~al.(2011)Nik\v{s}i\'{c}, Vretenar, and
  Ring}]{Niksic2011Prog.Part.Nucl.Phys.519}
\bibinfo{author}{T.~Nik\v{s}i\'{c}}, \bibinfo{author}{D.~Vretenar},
  \bibinfo{author}{P.~Ring}, \bibinfo{journal}{Prog. Part. Nucl. Phys.}
  \bibinfo{volume}{66} (\bibinfo{year}{2011}) \bibinfo{pages}{519--548}.
  \DOIprefix\doi{10.1016/j.ppnp.2011.01.055}.
\bibitem[{Meng(2015)}]{Meng2015}
\bibinfo{editor}{J.~Meng} (Ed.), \bibinfo{title}{Relativistic Density
  Functional for Nuclear Structure}, \bibinfo{publisher}{World Scientific},
  \bibinfo{year}{2015}.
\bibitem[{Zhao et~al.(2010)Zhao, Li, Yao, and Meng}]{Zhao2010Phys.Rev.C54319}
\bibinfo{author}{P.~W. Zhao}, \bibinfo{author}{Z.~P. Li},
  \bibinfo{author}{J.~M. Yao}, \bibinfo{author}{J.~Meng},
  \bibinfo{journal}{Phys. Rev. C} \bibinfo{volume}{82} (\bibinfo{year}{2010})
  \bibinfo{pages}{054319}. \DOIprefix\doi{10.1103/PhysRevC.82.054319}.
\bibitem[{Chen et~al.(2013)Chen, Zhang, Zhao, Jolos, and
  Meng}]{Chen2013Phys.Rev.C24314}
\bibinfo{author}{Q.~B. Chen}, \bibinfo{author}{S.~Q. Zhang},
  \bibinfo{author}{P.~W. Zhao}, \bibinfo{author}{R.~V. Jolos},
  \bibinfo{author}{J.~Meng}, \bibinfo{journal}{Phys. Rev. C}
  \bibinfo{volume}{87} (\bibinfo{year}{2013}) \bibinfo{pages}{024314}.
  \DOIprefix\doi{10.1103/PhysRevC.87.024314}.
\bibitem[{Chen et~al.(2016)Chen, Zhang, Zhao, Jolos, and
  Meng}]{Chen2016Phys.Rev.C44301}
\bibinfo{author}{Q.~B. Chen}, \bibinfo{author}{S.~Q. Zhang},
  \bibinfo{author}{P.~W. Zhao}, \bibinfo{author}{R.~V. Jolos},
  \bibinfo{author}{J.~Meng}, \bibinfo{journal}{Phys. Rev. C}
  \bibinfo{volume}{94} (\bibinfo{year}{2016}) \bibinfo{pages}{044301}.
  \DOIprefix\doi{10.1103/PhysRevC.94.044301}.
\bibitem[{Peng et~al.(2008)Peng, Meng, Ring, and
  Zhang}]{Peng2008Phys.Rev.C24313}
\bibinfo{author}{J.~Peng}, \bibinfo{author}{J.~Meng},
  \bibinfo{author}{P.~Ring}, \bibinfo{author}{S.~Q. Zhang},
  \bibinfo{journal}{Phys. Rev. C} \bibinfo{volume}{78} (\bibinfo{year}{2008})
  \bibinfo{pages}{024313}. \DOIprefix\doi{10.1103/PhysRevC.78.024313}.
\bibitem[{Bentz et~al.(1985)Bentz, Arima, Hyuga, Shimizu, and
  Yazaki}]{Bentz1985Nucl.Phys.A593}
\bibinfo{author}{W.~Bentz}, \bibinfo{author}{A.~Arima},
  \bibinfo{author}{H.~Hyuga}, \bibinfo{author}{K.~Shimizu},
  \bibinfo{author}{K.~Yazaki}, \bibinfo{journal}{Nucl. Phys. A}
  \bibinfo{volume}{436} (\bibinfo{year}{1985}) \bibinfo{pages}{593--620}.
  \DOIprefix\doi{10.1016/0375-9474(85)90550-0}.
\bibitem[{Arima(2011)}]{Arima2011SciChinaSerG-PhysMechAstron188}
\bibinfo{author}{A.~Arima}, \bibinfo{journal}{Sci. China Phys. Mech. Astron.}
  \bibinfo{volume}{54} (\bibinfo{year}{2011}) \bibinfo{pages}{188}.
  \DOIprefix\doi{10.1007/s11433-010-4224-6}.

\end{thebibliography}

\end{document}